\documentclass{andp2012}% no class options needed by now
\usepackage[english]{babel}

\usepackage{kantlipsum,cuted}

\usepackage[normalem]{ulem}
\usepackage{amsmath,amssymb}
\usepackage{multirow}
\usepackage{xcolor,soul}
\usepackage{srcltx}
\usepackage{hyperref,graphicx}

\title{Spectral structure and doublon dissociation in the two-particle non-Hermitian Hubbard model}
\author[S: Longhi]{Stefano Longhi \inst{1,}\footnote{Corresponding author\quad E-mail:~\textsf{stefano.longhi@polimi.it}}}
\address[1]{Dipartimento di Fisica, Politecnico di Milano,Piazza L. da Vinci 32, I-20133 Milano, Italy and IFISC (UIB-CSIC), Instituto de Fisica Interdisciplinar y Sistemas Complejos, E-07122 Palma de Mallorca, Spain}
\shortauthors{S. Longhi }
\begin{abstract}
\small
Strongly-correlated systems in non-Hermitian models are an emergent area of research. Here we consider a non-Hermitian Hubbard model, where the single-particle hopping amplitudes on the lattice are not reciprocal, and provide exact analytical results of the spectral structure in the two-particle sector of Hilbert space under different boundary conditions. 
{\color{black} The analysis unveils some interesting spectral and dynamical
effects of purely non-Hermitian nature and that deviate
from the usual scenario found in the single-particle regime.
Specifically, we predict} a spectral phase transition of the Mott-Hubbard band on the infinite lattice as the interaction energy is increased above a critical value, from an open to a closed loop in complex energy plane, and the dynamical dissociation of doublons, i.e. instability of two-particle bound states, in the bulk of the lattice, with a sudden revival of the doublon state when the two particles reach the lattice edge. {\color{black} Particle
dissociation observed in the bulk of the lattice is a clear manifestation
of non-Hermitian dynamics arising from the different
lifetimes of single-particle and two-particle states, whereas
the sudden revival of the doublon state at the boundaries is a
striking burst edge dynamical effect peculiar to non-Hermitian
systems with boundary-dependent energy spectra, here predicted
for the first time for correlated particles.}
\end{abstract}
\shortabstract
\begin{document}
\maketitle
% \noindent

\section{Introduction}
The Hubbard model \cite{r1} provides a powerful theoretical model in condensed matter physics and related fields, such as in ultracold atomic physics, that describes interacting electrons in a lattice. Since its introduction more than half a century ago, it has become a cornerstone of many studies, 
particularly in the field of strongly correlated systems (see e.g. \cite{r2,r3,r4} and references therein). The Hubbard model considers a lattice of sites and assumes that electrons can hop between neighboring lattice sites, with a Coulomb interaction occurring only when two electrons with opposite spins occupy the same site. The interplay between electron hopping and electron-electron interactions gives rise to a complex and rich phase diagram with various phases, such as unconventional superconductivity,
Mott insulators, and density-wave ordering,  depending on the values of the hopping amplitude, the interaction strength, and the electron filling \cite{r4}. Even in the simplest case of two interacting electrons with opposite spins, the Hubbard model 
displays an interesting physics, such as the formation of the Mott-Hubbard band describing doublons, i.e. pairs of bound repulsive electrons occupying the same lattice site \cite{r5,r6,r7,r8,r9,r10,r11,r12}.  
For sufficiently strong repulsion, doublons represent stable quasiparticles which undergo correlated tunneling on the lattice and cannot dissociate owing to energy conservation \cite{r13,r14}. \\ 
Recently, an enormous and increasing interest is being devoted toward the study of non-Hermitian models, where the Hamiltonian of the system is described by a non self-adjoint operator \cite{r15,r16,r17,r18,r19,r20,r21}. 
Effective non-Hermiticity originates
from exchanges of particles or energy with the external
environment {\color{black} and gives rise to unusual and counterintuitive
phenomena that are not present in Hermitian systems
\cite{r15,r16,r17,r18,r19,r20,r21,r22,r23,r24,r25,r26,r27,r28,r29,r30,r31,r32,r33,r34,r35,r36,r37,r38,r39,r40,r41,r42,r43,r44,r45,r46,r47,r48,r49,r50,r51,r52,r53,r54,r55}, such as novel topological phases and phase
transitions \cite{r16,r17,r18,r21,r22,r24,r25,r30,r32,r39}, exceptional
points and lines \cite{r20,r54,r55}, the non-Hermitian skin effect
\cite{r18,r22,r26,r27,r29,r34,r39,r54}, a generalized bulk-boundary
correspondence principle \cite{r18,r23,r26,r31,r35,r38,r43}, and
a wide variety of dynamical signatures and phenomena \cite{r34,r37,r50,r51,r52,r53}.
} Non-Hermitian extensions of the Hubbard and related models have been considered in some recent works \cite{r56,r57,r58,r59,r60,r61,r62,r63,r64,r65,r66,r67}. Different forms of the non-Hermitian Hubbard model have been introduced, which depend on the specific non-Hermitian terms included in the Hamiltonian. Generally, they can involve non-reciprocal hopping terms between lattice sites, complex on-site energies, or complex electron-electron interaction terms. 
Most of previous studies on non-Hermitian Hubbard models focussed on the
issues of non-Hermitian many-body localization in the presence of disorder, the interplay between correlation and the non-Hermitian skin effect, and many-body non-Hermitian topology. 
Studying the non-Hermitian Hubbard model involves analyzing the eigenvalues and eigenvectors of the Hamiltonian to understand the system energy levels and wavefunctions, their topological properties and novel phase transitions. The complexity of the analysis prevents rather generally to obtain exact analytical results, and large-scale numerical simulations are required even when dealing with few particles. The study of simple models allowing for exact analytical results is thus of main relevance and interest.
{\color{black} On the experimental
side, doublon dynamics has been demonstrated
in different physical platforms, including ultracold atoms
in optical lattices \cite{r5,r13,r14,r68,r69} and classical simulators
of two- or three-particle dynamics in Fock space based
on photonic \cite{r70} or topolectrical \cite{r71,r72} lattices with engineered
defects. The current advances in experimental
fabrication and control of synthetic matter, including the
ability to realize non-reciprocal hopping, make it possible
to realize few- body non-Hermitian Hubbard models in a
laboratory \cite{r72}, thus motivating the study of correlated particle
states and doublon dynamics in non-Hermitian
models.}

The simplest non-Hermitian version of the Hubbard model was introduced in Ref.\cite{r56} (see also \cite{r60,r61,r63}) and considers non-reciprocal single-particle hopping amplitudes arising from an imaginary gauge field. It can be regarded as a many-body generalization of the Hatano-Nelson model \cite{r73,r74,r75} for interacting particles, and can be thus referred to as the interacting Hatano-Nelson model. The Hatano-Nelson model provides a paradigmatic non-Hermitian model displaying a nontrivial point-gap topology and the non-Hermitian skin effect in the single-particle case \cite{r18}. One of the major results of single-particle band theory in models displaying the non-Hermitian skin effect is the strong dependence of the energy spectrum on the boundary conditions. Rather generally, under periodic boundary conditions (PBC) or in the infinite lattice limit (ILL) the energy spectrum is described by a set of closed loops in complex energy plane; under open boundary conditions (OBC) the energy spectrum shrinks into a set of open arcs embedded in the ILL energy loops; finally, under semi-infinite boundary conditions (SIBC) the energy spectrum is described by an area in complex energy plane whose contours are the ILL energy loops (see e.g. \cite{r39,r50}). As the spectral and  topological properties of the interacting Hatano-Nelson model under PBC have been numerically investigated in previous works \cite{r60,r61}, an open question is the dependence of the energy spectrum on the boundary conditions, either PBC, OBC or SIBC, and specifically how the general scenario found in the single-particle case \cite{r39} changes in the many-particle regime. Also, the  impact of non-reciprocal coupling on the dynamical behavior of correlated particles, such as correlated hopping and the stability of doublons, has been so far overlooked.\\
In this work we consider the interacting Hatano-Nelson model in the two-particle sector {\color{black} and unveil emergent
spectral and dynamical regimes arising from the interplay
between particle correlations and the non-Hermitian
gauge potential, which have been so far overlooked. By
a suitable extension of the Bethe Ansatz for the non-
Hermitian Hubbard model, we provide exact analytical
form of the energy spectrum for different types of boundary
conditions and unveil a new spectral phase transition
of theMott-Hubbard band of purely non-Hermitian nature
when the interaction strength is increased above a
critical value. Finally, we predict a novel dynamical regime
of correlated particles without anyHermitian counterpart,
namely doublon dissociation in the bulk of the lattice and
sudden doublon resurgence at the lattice edges. Particle
dissociation observed in the bulk of the lattice arises
from the longer lifetime of single-particle states over two-particle
states, and it is thus a very distinct phenomenon
that particle delocalization observed in the single-particle
Hatano-Nelson model with disorder \cite{r73,r74,r75}. On the other
hand, the sudden resurgence of the doublon state at
the lattice boundaries is a striking boundary-induced dynamical
phenomenon peculiar to non-Hermitian systems
displaying the non-Hermitian skin effect \cite{r51,r52}, here
demonstrated for the first time for correlated particle
states.}

%Overall, non-Hermitian physics provides a new perspective for under standing and describing complex systems that go beyond the traditional Hermitian framework of quantum mechanics. It offers intriguing opportunities for exploring novel phenomena and designing innovative devices in various areas of physics and engineering.
%Overall, non-Hermitian physics provides a framework for exploring and understanding physical systems that go beyond the traditional Hermitian formalism. It opens up new avenues for research and has the potential to lead to the development of novel technologies and a deeper understanding of fundamental physical principles.
%The study of non-Hermitian topological phases is still an emerging field, and researchers are actively exploring their fundamental properties, potential applications, and the interplay with other areas of physics, such as quantum information and quantum simulation. The understanding of non-Hermitian topological phases has the potential to uncover new physical phenomena and facilitate the development of novel devices and technologies based on their unique properties.

\section{The non-Hermitian Hubbard model for two interacting particles}
The starting point of our analysis is provided by a non-Hermitian extension of the Hubbard model \cite{r56,r61}, also referred to as the interacting Hatano-Nelson model, which describes the hopping dynamics of interacting fermionic particles in a one-dimensional tight-binding lattice subjected to an imaginary gauge field. We indicate by $J>0$ the single-particle hopping amplitude between adjacent sites in the lattice, by $h \geq 0$ the imaginary gauge field, and by $U$  the on-site interaction energy of fermions with opposite spins ($U>0$ for a repulsive interaction). 
The effective non-Hermitian Hubbard Hamiltonian of the system reads \cite{r56,r61}
\begin{eqnarray}
 \hat{H}  & = &   - \sum_{l, \sigma} \left( J \exp(h) \hat{a}^{\dag}_{l, \sigma}  \hat{a}_{l+1, \sigma}+ J \exp(-h) \hat{a}^{\dag}_{l+1, \sigma} \hat{a}^{}_{l, \sigma}   \right) \nonumber \\
 &   + &   U \sum_{l} \hat{n}_{l, \uparrow}  \hat{n}_{l, \downarrow}  
\end{eqnarray}
where $\hat{a}_{l, \sigma}$, $\hat{a}^{\dag}_{l, \sigma}$ are the annihilation and creation operators of fermions with spin $\sigma= \uparrow, \downarrow$
 at lattice site $l$, $\hat{n}_{l, \sigma}= \hat{a}^{\dag}_{l, \sigma} \hat{a}_{l, \sigma}$ is the particle-number operator, and $h$ is the imaginary gauge field (magnetic flux).
 In the limit $U=0$, the model reduces to the many-particle non-interacting Hatano-Nelson model, whereas for $h=0$ it reduces to the standard (Hermitian) Hubbard model. 
 For pure states and considering open-system dynamics conditioned on
measurement outcomes such that the quantum evolution corresponds to the null-jump process, the state vector $| \psi(t) \rangle$ 
of the system at time $t$ reads (see e.g. \cite{r57,r76,r77,r78,r79})
\begin{equation}
| \psi(t) \rangle= \frac {\exp(-i \hat{H} t) | \psi(0) \rangle} {\| \exp(-i \hat{H} t) | \psi(0) \rangle \|} \equiv \frac{ | \Phi(t) \rangle}{ \| | \Phi(t) \rangle \|}
\end{equation}
where we have set $| \Phi(t) \rangle \equiv \exp(-iH t) | \psi(0) \rangle$. Basically, at each time interval $dt$ the state vector evolves according to the Schr\^odinger equation with an effective non-Hermitian Hamiltonian $\hat{H}$, followed by a normalization of the wave function, without undergoing any quantum jump \cite{r57,r76,r77,r78,r79}.\\
In the single-particle sector of Hilbert space, the Hamiltonian (1) describes the well-known disorder-free Hatano-Nelson model \cite{r73,r74,r75}, which provides a paradigmatic model displaying the non-Hermitian skin effect, a nontrivial point-gap topology and a strong dependence of its energy spectrum on the boundary conditions (see e.g. \cite{r16,r18,r39}).
  In this work we will focus our analysis by considering two fermions with opposite spins.
  In this case, we can expand $| \Phi(t) \rangle$ in Fock space according to
\begin{equation}
| \Phi(t) \rangle = \sum_{n,m}  \phi_{n,m}(t) \hat{a}^{\dag}_{n, \uparrow} \hat{a}^{\dag}_{m, \downarrow} | 0 \rangle,
\end{equation}
where $|\phi_{n,m}(t)|^2$ is the (non-normalized) probability that the two electrons with spin $\uparrow$ and $\downarrow$ occupy the two sites $n$ and $m$ of the lattice, respectively. The evolution equations of the amplitudes $\phi_{n,m}(t)$ read
\begin{eqnarray}
i \frac{d \phi_{n,m}}{dt} &  = & - J \exp(h) \left( \phi_{n+1,m}+\phi_{n,m+1} \right) \\
& - & J \exp(-h) \left(\phi_{n-1,m}+\phi_{n,m-1} \right) +U \phi_{n,m} \delta_{n,m}. \nonumber
\end{eqnarray}

\section{Energy spectrum}
Let us first briefly remind the spectral properties in the single-particle sector of Hilbert space, where the non-Hermitian Hubbard model reduces to the clean (disorder-free) Hatano-Nelson model. 
The results are summarized in Fig.1 (see e.g. \cite{
r16,r18,r39} for details).
The energy spectrum $\sigma(H_{ILL})$ on the infinite lattice (or equivalently under PBC) is given by
\[
E=-2J \cos(k-ih)
\]
where $k$ is the Bloch wave number,  that can take real values and varies in the range $-\pi \leq k \leq \pi$. Clearly, the energy spectrum under PBC describes a closed loop (an ellipse) in complex energy plane (solid red curve in Fig.1). The corresponding eigenfunctions are plane waves with a real wave number $k$. Under OBC, the energy spectrum $\sigma(H_{OBC})$ is entirely real, independent of the imaginary gauge field $h$, and reads (green solid curve in Fig.1)
\[
E=-2J \cos(k)
\]
($-\pi \leq k \leq \pi$); the corresponding eigenfunctions are squeezed toward one edge of the lattice (non-Hermitian skin effect). Finally, under SIBC the energy spectrum $\sigma(H_{SIBC})$ reads (shaded blue area in Fig.1)
\[
E=-2J \cos(k-ih)
\]
 where now the wave number $k$ is complex satisfying the constraint $0 \leq {\rm Im}(k) \leq 2h$. In complex energy plane, the SIBC energy spectrum describes the entire area (ellipse) enclosed by the PBC energy loop, and the corresponding wave functions are extended when $k$ is real and squeezed toward the edge of the semi-infinite lattice when $k$ is complex. We mention that each SIBC edge eigenstate, with eigenenergy $E$ internal to the PBC energy loop, is predicted from a bulk-boundary correspondence principle \cite{r16,r39} and should not be regarded as a mere mathematical object, because it can be selectively excited and it is thus of physical relevance \cite{r50,r80}.
   
 \begin{figure}
  \includegraphics[width=8cm]{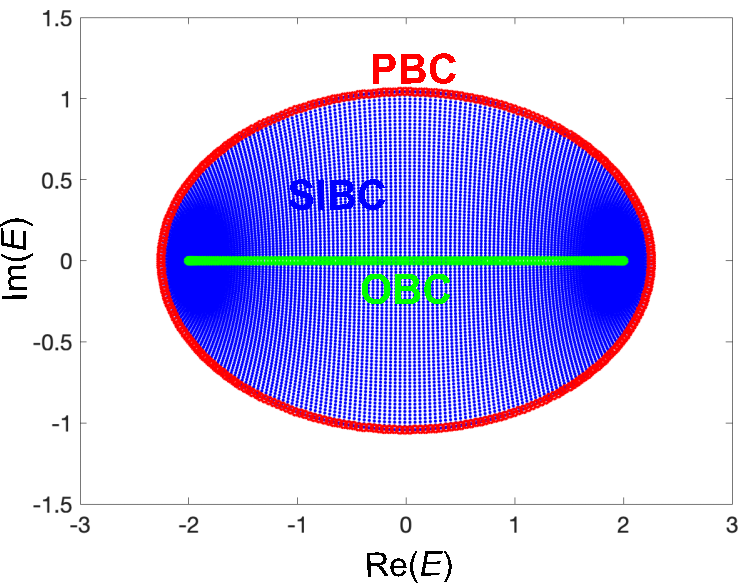}
  \caption{Energy spectrum $\sigma(H)$ in complex energy plane of the single-particle non-Hermitian Hubbard model (Hatano-Nelson model) under different boundary conditions. The outer red loop (ellipse) is the energy spectrum $\sigma(H_{PBC})$ (or $\sigma(H_{ILL})$) under PBC, the straight segment on the real energy axis is the OBC energy spectrum $\sigma(H_{OBC})$, whereas the shaded blue area is the SIBC energy spectrum $\sigma(H_{SIBC})$. Parameter values are $J=1$ and $h=0.5$.}
\end{figure}
In the two-particle sector of Hilbert space, the energy spectrum $E$ and corresponding eigenstates $u_{n,m}$ of the non-Hermitian Hubbard Hamiltonian are obtained by solving the spectral problem
\begin{eqnarray}
E u_{n,m} &  = & - J \exp(h) \left( u_{n+1,m}+u_{n,m+1} \right) \\
& - & J \exp(-h) \left(u_{n-1,m}+u_{n,m-1} \right) +U u_{n,m} \delta_{n,m} \nonumber
\end{eqnarray}
with suitable boundary conditions. The eigenstates can be classified as symmetric or antisymmetric for particle exchange, i.e. $u_{n,m}=u_{m,n}$ or  $u_{n,m}=-u_{m,n}$.
%In fact, it can be readily shown that, if $u_{n,m}$ is an eigenfunction to Eq.(5) with eigenenergy $E$, then $w_{n,m}=u_{m,n}$ is an eigenfunction as well with the same eigenenergy. Hence, if $u_{n,m} \neq  \pm u_{m,n}$, for each eigenenergy $E$ we can construct the couple of linearly-independent eigenfunctions $(u_{n,m} \pm u_{m,n}) / \sqrt{2}$ which are either symmetric or antisymmetric for particle exchange.
 Clearly, since for an antisymmetric eigenfunction one has $u_{n,n}=0$, the interaction term $U$ does not influence the energy spectrum.  Hence, for antisymmetric states the spectrum can be simply retrieved from the spectrum of the single particle problem. Conversely, the interaction term is important for symmetric states. Therefore, in the following we will limit to consider the energy spectrum arising from the wave functions which are symmetric under particle exchange, i.e. we impose the further condition $u_{n,m}=u_{m,n}$ to Eq.(5).

The energy spectrum in the Hermitian limit $h=0$ is exactly solvable using the Bethe-Ansatz method \cite{r3}. Usually, the solution is searched by assuming a lattice of finite size $L$ with either PBC or OBC, which provide suitable quantization conditions of wave numbers via the  Lieb-Wu equations \cite{r3}. However, a much simpler situation occurs when considering an infinite lattice and the solutions on the infinite interval \cite{r3,r6,r81,r82}, which do not require quantization of the wave numbers. In this case,  the energy spectrum is absolutely continuous and composed by two branches, corresponding to (i) the scattering states of asymptotically free particles ($E_1$-branch), and (ii) interaction-bound dimer states (doublons; $E_2$-branch) \cite{r3,r6}. The energy spectrum of scattering states ($E_1$-branch) is simply given by
\begin{equation}
E_{1}(k_1,k_2)=-2J \cos k_1-2J \cos k_2
\end{equation}
where $k_1$, $k_2$ are arbitrary real wave numbers (quasi momenta) of the two asympotically-free fermions. This spectrum is precisely the one found for two non-interacting particles; when considering both symmetric and antisymmetric states, this implies that the above energies are doubly degenerate (the degeneracy is lifted for a finite size $L$ of the lattice \cite{r3}). For the symmetric states, we have the additional  energy spectrum describing the bound-particle state (doublon) branch ($E_2$-branch), given by
\begin{equation}
E_{2}(q)= \sqrt{U^2+ 16 J^2 \cos^2 (q)}
\end{equation}
 \begin{figure}
  \includegraphics[width=8cm]{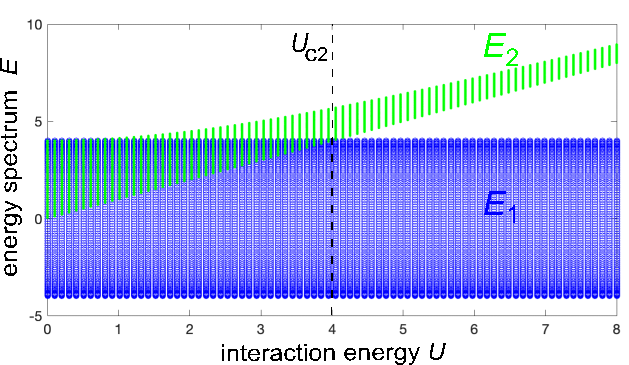}
  \caption{Energy spectrum $\sigma(H)$ of the two-particle Hubbard model in the Hermitian limit $h=0$ versus the interaction energy $U$ for $J=1$. The spectrum comprises two branches, $E_1$ and $E_2$ branches, corresponding to asymptotically two-particle free scattering states and bound two-particle states (doublons), respectively. An energy gap appears as $U$ is increased above the critical value $U_{c2}=4J$. In the non-Hermitian regime $h>0$, the same spectrum is obtained under OBC.}
\end{figure}
 where $q=(k_1+k_2)/2$ is the real wave number of the particle center of mass. The latter branch corresponds to the well-known Mott-Hubbard band, with the formation of a gap when the interaction energy $U$ is larger than the critical value $U_{c2}=4J$; see Fig.2. Clearly, in the Hermitian limit the above results can be obtained from the finite lattice model, with either PBC or OBC, in the thermodynamic limit $L \rightarrow \infty$.\\
 The situation drastically changes when we consider the non-Hermitian Hubbard model with a non-vanishing imaginary gauge field $h>0$, which makes the energy spectrum $\sigma(H)$ strongly dependent on the boundary conditions even in the absence of particle interaction (see e.g. \cite{r16,r18,r28,r29,r39}). In the following analysis, we will provide exact results of the energy spectrum $\sigma(H)$ for $U>0$ in the two-particle sector of Hilbert space by considering three different boundary conditions:\\ 
  (i) OBC: a finite lattice comprising $L$ sites with OBC in the large $L$ limit. In this case the spectral problem (5) is defined on the domain $n,m=1,2,...,L$ with the boundary conditions
  \begin{equation}
  u_{0,m}=u_{n,0}=u_{L+1,m}=u_{n,L+1}=0.
  \end{equation}
 (ii) ILL: the infinitely-extended lattice, where the lattice indices $n$ and $m$ vary over the entire infinite interval $(-\infty, \infty)$. In this case the spectral problem (5) is supplemented by the condition that $|u_{n,m}|$ remains bounded as $n,m \rightarrow \pm \infty$, i.e.
 \begin{equation}
 {\rm Sup}{\lim}_{n,m \rightarrow \pm \infty} |u_{n,m}| < \infty.
 \end{equation}
 We mention that the ILL boundary conditions greatly simplify the analysis than considering more common PBC on a finite lattice of size $L$, in the large $L$ limit. We will check by numerical simulations that, as one would expect by considering the $U=0$ limit, the PBC energy spectrum in the large $L$ limit is reproduced by the ILL energy spectrum, which can be calculated in an exact form.\\
 (iii) SIBC: a semi-infinite lattice. In this case the spectral problem (5) is defined over the semi-plane $n,m=1,2,3,...$ with the boundary conditions
  \begin{equation}
  u_{0,m}=u_{n,0}=0
  \end{equation}
and
 \begin{equation}
 {\rm Sup}{\lim}_{n,m \rightarrow + \infty} |u_{n,m}| < \infty.
 \end{equation}
\subsection{Energy spectrum under open boundary conditions}
Assuming OBC, the spectral problem defined by Eqs.(5) and (8) can be readily solved by introduction of a non-unitary transformation of the wave function, which reduces the analysis to the ordinary (Hermitian) Hubbard model. After letting
\begin{equation}
u_{n,m}=w_{n,m} \exp(-hn-hm)
\end{equation}
from Eq.(5) one obtains
\begin{eqnarray}
E w_{n,m} &  = & - J  \left( w_{n+1,m}+w_{n,m+1} \right. \\
& - & \left. w_{n-1,m}+w_{n,m-1} \right) +U w_{n,m} \delta_{n,m} \nonumber
\end{eqnarray}
 which corresponds to the spectral problem of the Hermitian Hubbard model in the two-particle sector under OBC. In the large $L$ limit, the energy spectrum remains entirely real and is thus given by Eqs.(6) and (7) and depicted in Fig.2. In other words, like in the Hatano-Nelson model under OBC the energy spectrum is not modified by the application of the imaginary gauge field $h$, however according to Eq.(12) the eigenstates are squeezed toward the corner $n=m=0$, corresponding to the non-Hermitian skin effect for the two interacting particles.  
 \subsection{Energy spectrum on the infinite lattice}
 In the infinite lattice, we can solve the spectral problem defined by Eqs.(5) and (9) using the same procedure as in the Hermitian Hubbard model, decomposing the wave function in terms of the center of mass and relative spatial coordinates (see e.g. \cite{r6}). After letting
 \begin{equation}
 u_{n,m}=\Phi_{n-m} \exp[iq(n+m)]
 \end{equation}
 for $m \leq n$, and $u_{n,m}=u_{m,n}$ for $m \geq n$, from Eq.(5) one obtains the following difference equation for the amplitudes $\Phi_l$, with $l=n-m$ the relative spatial coordinate between the two fermions
\begin{equation}
E \Phi_l= -2 J \sigma (\Phi_{l+1}+\Phi_{l-1})+U \delta_{l,0} \Phi_0
\end{equation} 
 and $\Phi_l=\Phi_{-l}$. In the above equation we have set 
 \begin{equation}
 \sigma \equiv  \cos(q-ih). 
 \end{equation}
 and the parameter $q=(k_1+k_2)/2$ in Eq.(14) plays the role of the quasi-momentum of the center of mass of the two fermions. Clearly, the asymptotic condition (9) requires $q$ to be a real number. 
 The eigenstates and corresponding energy spectra of Eq.(15), which are bounded as $l \rightarrow \pm \infty$, are calculated in Appendix A and consists of two branches ($E_1$- and $E_2$-energy branches)(i) a set of scattered states with asymptotic behavior $\Phi_l \sim \cos (Q l)$ as $l \rightarrow \infty$, where $Q=(k_1-k_2)/2$ is an arbitrary real parameter which physically describes the quasi-momentum of the relative particle motion; and (ii) a set of bound states, namely $\Phi_l= \exp (-\mu |l| ) $ ($ l \neq 0$) with $\mu>0$, corresponding to the bounded two-particle state (doublon) branch. The dispersion curve of the scattered two-particle states ($E_1$-energy branch) is given by
 \begin{equation}
 E_1=-4 J \cos (q-ih) \cos Q=-2J \cos (k_1-ih)-2 J \cos(k_2-ih)
 \end{equation}
 whereas the dispersion curve of the bound two-particle states  ($E_2$-energy branch) reads
 \begin{equation}
 E_2=\sqrt{U^2+16 J^2 \cos^2(q-ih) }
 \end{equation}
  \begin{figure*}
  \includegraphics[width=17cm]{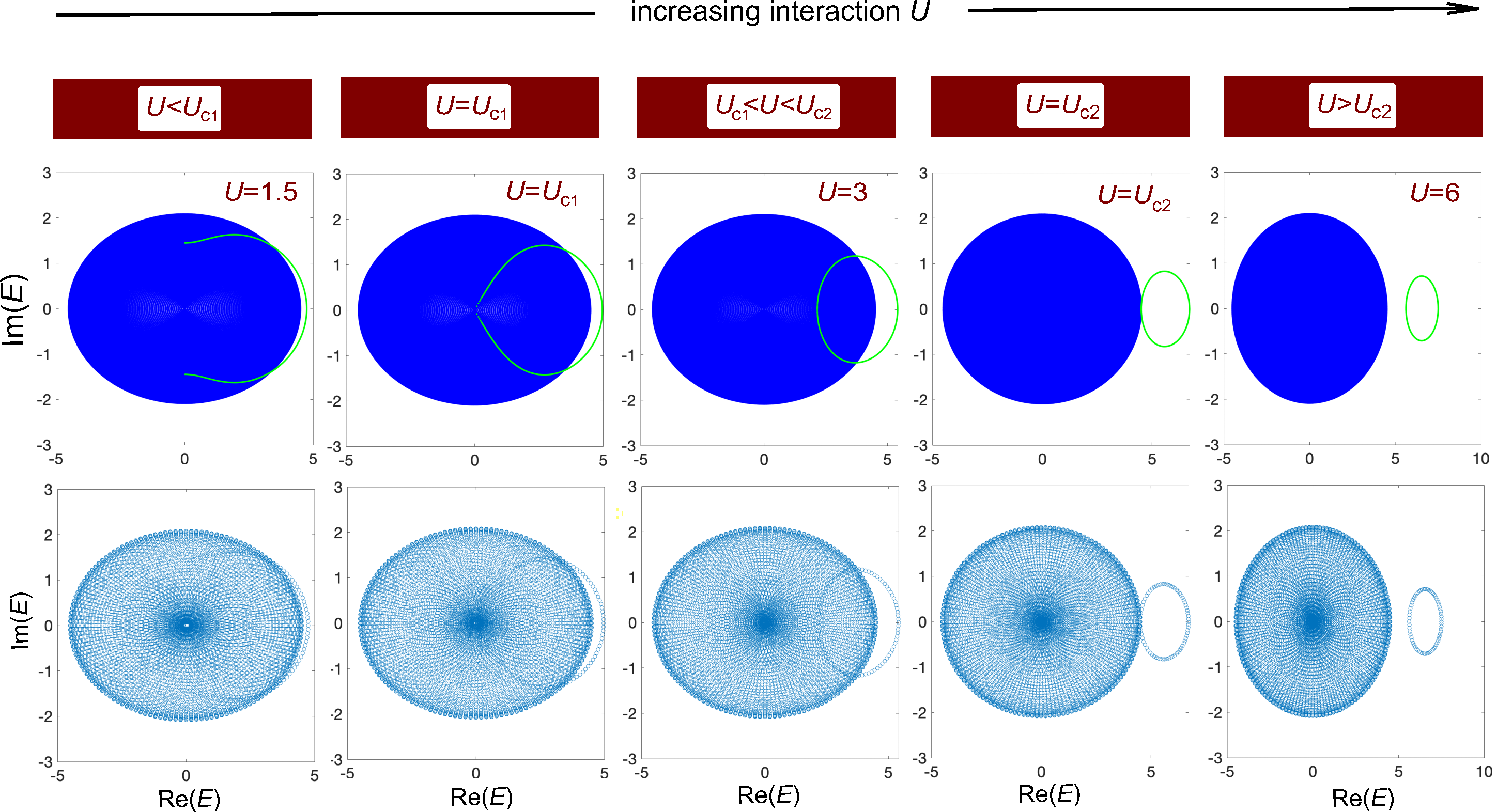}
  \caption{Energy spectrum $\sigma(H_{ILL})$ of the two-particle non-Hermitian Hubbard model on an infinite lattice for increasing interaction energy $U$. Parameter values are $J=1$ and $h=0.5$. The critical values of the interaction energy $U_{c1}$ and $U_{c2}$ are given by Eqs.(19) and (20), and for the specific parameter values read $U_{c1} \simeq 2.0844$ and $U_{c2} \simeq 4.9688$. The upper row shows the energy spectra ($E_1$- and $E_2$-energy branches) as obtained by the exact analysis. The shaded area, independent of $U$, describes the energy spectrum of the scattered two-particle states [Eq.(17), $E_1$ branch], whereas the solid light curve describes the doublon (Mott-Hubbard) energy band [Eq.(18), $E_2$ branch]. For $U<U_{c1}$ ($U>U_{c1}$) the Mott-Hubbard band describes an open (closed) loop in complex energy plane. For $U>U_{c2}$ a line gap separates the two bands. The lower row shows the energy spectra as computed by numerical diagonalization of the matrix Hamiltonian in a finite lattice comprising $L=70$ sites and assuming PBC.}
\end{figure*}
Note that the two above energy branches can be obtained from the corresponding dispersion curves in the Hermitian limit [Eqs.(6) and (7)] after the substitution $k_{1,2} \rightarrow k_{1,2}-ih$ and $q \rightarrow q-ih$.
Typical examples of the energy spectra for increasing values of the interaction energy $U$, as given by the analytical expression Eqs.(17) and (18), are shown in the upper row of Fig.3. For comparison, the energy spectra numerically computed in a finite lattice of size $L=70$ by matrix diagonalization and applying PBC are depicted in the lower row of Fig.3, indicating that the analytical results well approximate the PBC energy spectra in the large $L$ limit. Note that the energy spectra associated to the two-particle scattered states cover an entire area of the complex energy plane. Clearly, this result also occurs for non-interacting particles and is due to the fact that we are dealing with two particles and the energy dispersion curve $E_1$ depends on two real parameters $k_1$ and $k_2$. The fact that the two-particle energy spectrum is described by an area in complex plane is the clear signature of the non-Hermitian skin effect \cite{r54}. On the other hand, the energy spectrum associated to the two-particle bound states (doublon) is described by a curve in complex energy plane, since in this case the dispersion curve of the $E_2$-energy branch depends only on one real parameter $q=(k_1+k_2)/2$. It can be readily shown that such a curve describes an open arc when $U<U_{c1}$, whereas it describes a closed loop when $U>U_{c1}$, where the critical interaction energy $U_{c1}$ is given by
\begin{equation}
U_{c1}=4 J \sinh h.
\end{equation}
Therefore, the doublon (Mott-Hubbard) $E_2$-energy branch shows a first spectral phase transition as $U$ is increased above $U_{c1}$. 
{\color{black} The fact that for $U<U_{c1}$ the energy spectrum
of the Mott-Hubbard band describes an open (rather than
a closed) loop is a rather unexpected and remarkable result,
since in single-particle models displaying the non-Hermitian skin effect the energy spectra are described by
open loops with a nontrivial point-gap topology.}

At a larger interaction strength $U_{c2}$, given by
\begin{equation}
U_{c2}= 4 J \sqrt{\cosh^2 h+ \sinh^2 h}
\end{equation}
 the Mott-Hubbard spectral loop $E_2$ fully detaches from the scattered two-particle energy spectrum $E_1$, which are separated one other by a line gap. This behavior is clearly illustrated in the upper row of Fig.3. While the detachment of the Mott-Hubbard band as the interaction energy $U$ is increased above the critical value $U_{c2}$ is similar to the behavior found in the ordinary (Hermitian) Hubbard model (see Fig.2), the spectral phase transition of the Mott-Hubbard band occurring at the lower interaction energy $U_{c1}$, from an open to a closed loop in the complex energy plane, is a purely non-Hermitian phenomenon without any counterpart in the Hermitian Hubbard model. 
 
 \subsection{Energy spectrum on the semi-infinite lattice}
 The energy spectrum on the semi-infinite lattice $\sigma(H_{SIBC})$ requires to solve the spectral problem defined by Eqs.(5), (10) and (11). As compared to the infinite lattice case considered in the previous subsection, the analysis is more involved and the technical details are given in the Appendix B. The starting point is provided by a Bethe-like Ansatz \cite{r3,r83} of the wave function $u_{n,m}$, given by the superposition of eight plane waves obtained by the set of permutations of  the complex wave numbers $\pm k_1$ and $\pm k_2$ for the two particles, suitably modified to account for a non-vanishing imaginary gauge field $h$.  
 As shown in the Appendix B and illustrated in Fig.4, the energy spectrum consists of two branches. The first branch ($E_1$-energy branch) is defined by the dispersion relation 
 \begin{equation}
 E_1=-2J \cos (k_1-ih)-2J \cos(k_2-ih)
 \end{equation}
 \begin{figure*}
  \includegraphics[width=17cm]{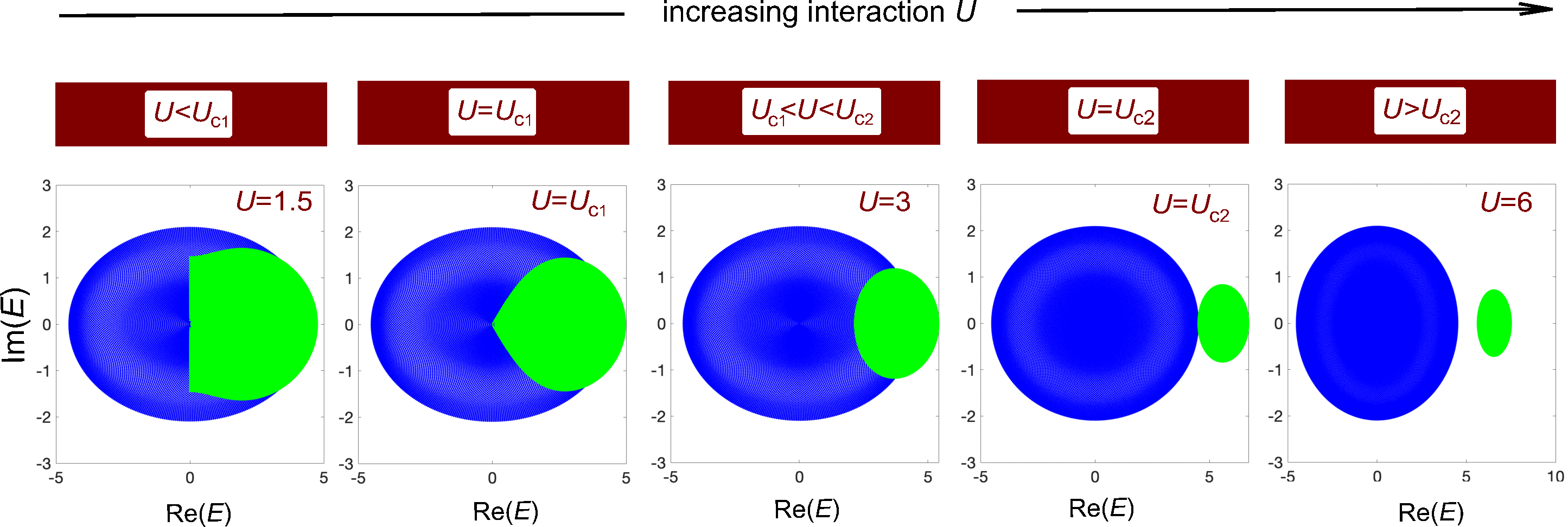}
  \caption{Energy spectrum $\sigma(H_{SIBC})$ of the two-particle non-Hermitian Hubbard model on an semi-infinite lattice for increasing interaction energy $U$. Parameter values are as in Fig.3 ($J=1$, $h=0.5$). The dark shaded area, independent of $U$, describes the $E_1$ energy branch, whereas the light shaded area describes the spectrum of the doublon eigenstates ($E_2$ branch). The critical values $U_{c1}$ and $U_{c2}$ are given by Eqs.(19) and (20). }
\end{figure*}
 which is formally the same as the dispersion curve of two-particle scattered states found in the infinite lattice case [Eq.(17)]. However, under SIBC the wave numbers $k_1$ and $k_2$ can be complex and should satisfy the minimal constraint
\begin{equation}
0 \leq {\rm Im} (k_{1,2} ) \leq 2h.
\end{equation}
By restricting $k_1$ and $k_2$ to be real, i.e. for ${\rm Im} (k_{1,2})=0$, we retrieve the $E_1$-energy branch on the infinite lattice derived in Sec.3.2 and corresponding to scattered two-particle states. Remarkably, under SIBC even though we allow the imaginary parts of $k_1$ and $k_2$ to be non vanishing and to vary in the range constrained by Eq.(22), the $E_1$-branch of the energy spectrum does not change and describes the same area in complex energy plane as in the infinite lattice case (see the shaded dark areas in Figs.3 and 4, which are the same and independent of $U$). However, when the imaginary parts of $k_1$ and $k_2$ do not vanish, the wave functions are not anymore extended (scattered states), rather they become squeezed toward the edge $n=m=1$ (two-particle non-Hermitian skin effect). Finally, we note that by restricting ${\rm Im} (k_{1,2})=h$, the $E_1$ branch of the SIBC energy spectrum reduces to the $E_1$-branch of the OBC energy spectrum [Eq.(6)].\\
The second branch of the SIBC energy spectrum, the $E_2$-branch, is defined by the dispersion relation
 \begin{equation}
 E_2=\sqrt{U^2+16 J^2 \cos^2(q-ih) }
 \end{equation}
which is formally the same as the $E_2$ branch on the infinite lattice [Eq.(18)]. However, under SIBC the wave number $q=(k_1+k_2)/2$ is allowed to have a non-vanishing imaginary part, satisfying the minimal constraint
\begin{equation}
0 \leq {\rm Im} (q) \leq  h.
\end{equation}  
 As a consequence, while the $E_2$-energy branch in the infinite lattice is described by an open arc or a closed loop (Fig.3), under SIBC the $E_2$-energy branch describes an area in complex energy plane, as shown in Fig.4. Since the Mott-Hubbard $E_2$-energy branch emerges only when the interaction energy $U$ is non-vanishing, we can conclude that while in the interaction-free regime $U=0$ the two energy spectra $\sigma(H_{ILL})$ and $\sigma(H_{SIBB})$
 are the same, particle interaction lifts such degeneracy and the two spectra differ one another under the two different boundary conditions. 
\begin{figure*}
  \includegraphics[width=17cm]{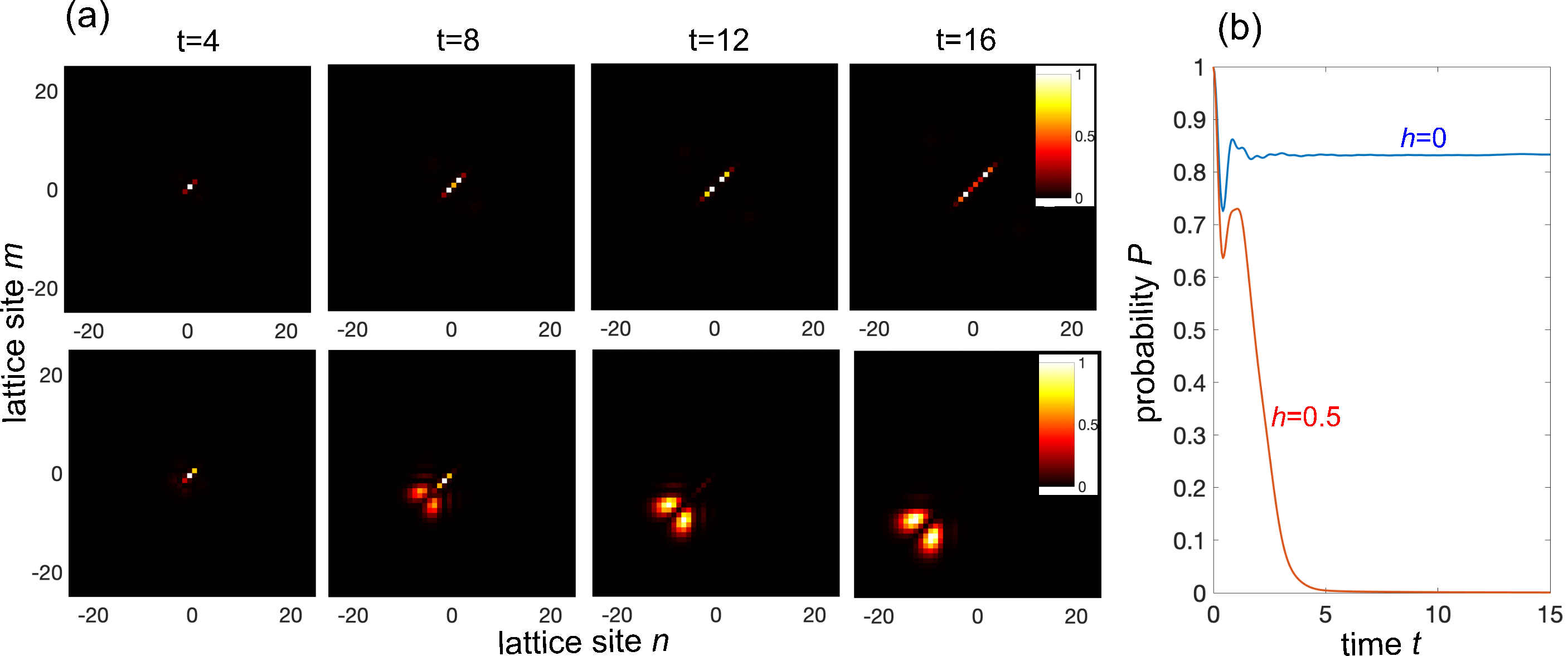}
  \caption{Dynamical behavior of the two-particle non-Hermitian Hubbard model. (a) Temporal evolution of the two-particle occupation probabilities $|\psi_{n,m}(t)|^2$ on the lattice as obtained by numerical integration of Eqs.(4) for the initial condition $\psi_{n,m}(0)=\delta_{n,0} \delta_{m,0}$, corresponding to the two fermions occupying the same lattice site $n=0$. Parameter values are $J=1$, $U=6$, and $h=0$ (Hermitian limit, upper panels) and $h=0.5$ (lower panels). 
  {\color{black} The occupation probabilities are plotted using a pseudocolor map and,
at each time instant, they are normalized to the peak value.} The lattice size is large enough ($L=51$) to avoid edge effects up to the largest observation time $t=16$. (b) Corresponding temporal evolution of the same-site occupation probability $P(t)$. }
\end{figure*}

\section{Two-particle dynamics: non-Hermitian induced doublon dissociation and burst edge revival}
One of the major results of the Hubbard model in the Hermitian limit is 
the formation of stable pairs of bound
particles occupying the same lattice site, dubbed doublons \cite{r5,r6,r7,r8,r9,r10,r11,r12}.   
For sufficiently strong interactions, isolated doublons represent stable quasiparticles
which undergo correlated tunneling on the lattice \cite{r13,r14}. 
Doublon dynamics has been experimentally observed using different platforms, such as ultracold atoms \cite{
r5,r13,r14,r68} and classical systems \cite{r70,r71}. 
A natural question is whether two strongly-interacting particles remain bounded and undergo correlated tunneling on the lattice when we apply the non-Hermitian gauge field. 

The {\em bulk} dynamical properties of the two-particle system, i.e. far from lattice edges,  can be captured by expanding the state vector of the system on the basis of the 
eigenstates of the two-particle Hamiltonian on the infinite lattice \cite{r34,r76,r84},  {\color{black}with amplitudes given in
terms of the scalar products of the initial state with the
left eigenstates of the Hamiltonian. In our case, left and
right eigenstates are just obtained one another by a sign
flip of $h$, i.e. by the transformation $h  \rightarrow -h$.}  The resulting asymptotic dynamics is dominated by the eigenstates { \color{black} with the longer lifetime (for decaying states) or the larger growth rate (for unstable growing states) \cite{r34,r76,r84,r85}}.  
The energy spectra shown in Fig.3 clearly show that the dominant eigenstates, with the largest {\color{black} growth rate}, belong to the $E_1$-energy branch and thus they describe dissociated particle states, i.e. fermions that do not occupy the same lattice site. 
This means that, even for a strong interaction energy, two fermions with opposite spins initially placed at the same lattice site and forming a doublon state do not remain bounded anylonger and they dissociate in the long time evolution. In other words, non-Hermiticity makes doublons unstable quasi-particle states. This result is clearly illustrated in Fig.5. The figure shows the numerically-computed temporal evolution of the two-particle probability occupation probabilities $|\psi_{n,m}(t)|^2$ [panel (a)] and of the probability $P(t)$ that the two fermions at time $t$ occupy the same lattice site [panel(b)], i.e.
\begin{equation}
P(t)= \sum_n | \psi_{n,n}(t)|^2.
\end{equation}
 \begin{figure}
  \includegraphics[width=8cm]{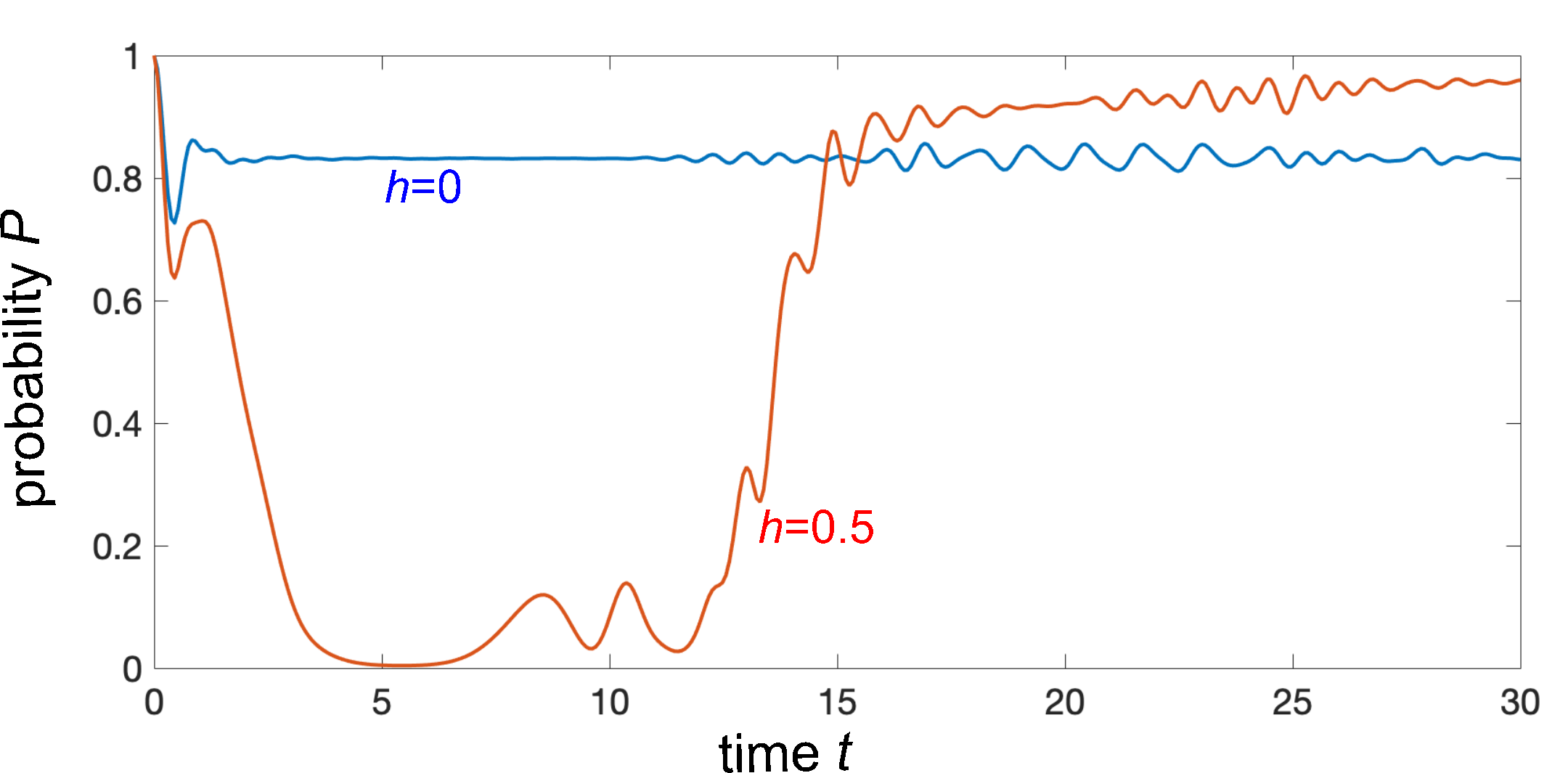}
  \caption{Edge burst revival of the same-site occupation probability $P(t)$ due to boundary effects. The figure shows the behavior of the probability $P(t)$ for the same parameter values as in Fig.5(b), but on a finite lattice comprising $L=31$ sites. When the two particles reach the lattice boundary, a sudden revival of $P(t)$ is observed in the non-Hermitian regime.}
\end{figure}
At initial time, the two fermions are placed at the same site $n=0$ of the lattice. While in the Hermitian limit the probability $P(t)$ reaches a stationary and large value, indicating the stability of the doublon state, in the non-Hermitian regime the probability $P(t)$ rapidly decays toward zero, indicating particle dissociation. 
{\color{black}
It should be
mentioned that particle dissociation shown in Fig.5 and
induced by the imaginary gauge field is a very distinct
phenomenon than non-Hermitian delocalization found
in the Hatano-Nelson model with lattice disorder \cite{r73,r74,r75}.
Delocalization concerns with the extended nature of either
single-particle or two-particle eigenstates; in presence
of disorder, single- and two-particle states that are
Anderson localized undergo a delocalization transition
when an imaginary gauge field is applied. On the other
hand, dissociation concerns with the lifetime (rather than
localized or extended nature) of the two-particle wave
functions. In our model we do not have any disorder in
the system and so all states are delocalized, even when
the imaginary gauge field $h$ vanishes: so dynamical delocalization
is observed for both single- and two-particle
states. Dissociation of doublon states observed in Fig.5
is not related to delocalization, rather it originates physically 
from the larger growth rates of asymptotically-free two-particle
states than those of two-particle bound states. This is because, when the two particles are sticked together, correlated tunneling occurs and spreading of bound states in the lattice is slower than that of single-particle states \cite{r86}. Correspondingly, when the left/right hopping rates are non-reciprocal, two-particle bound state wave packets (doublons) experience a lower amplification than asymptotically-free two-particle wave packets when they drift and spread along the bulk of the lattice, resulting in the effective dissociation of the particles under continuous measurements of the system. It should be mentioned that with additional non-Hermitian terms in  the Hubbard model, for
example by including an additional two-particle
gain term by replacing $U$ with $U+iG$ ($G>0$) in the last
term of Eq.(1), the growth rates of doublon states could become
larger than those of asymptotically-free two-particle
states, so that in such a regime doublons do not dissociate, the two particles remain sticked together, yet the bound state  wave packet spreads and delocalizes in the lattice.

The doublon dissociation shown in Fig.5 is clearly
a bulk dynamical phenomenon. Remarkably, when the
particles reach the boundaries of the lattice a burst revival
of $P(t )$, corresponding to a sudden resurgence of
the doublon state, is observed, as shown in Fig.6. Such
an edge burst revivial is a striking boundary-induced
dynamical phenomenon peculiar to non-Hermitian systems
displaying the non-Hermitian skin effect, and extend
to correlated particle systems edge burst dynamical phenomena
predicted and observed in single-particle non-Hermitian models \cite{r51,r52}. From a physical viewpoint, the
edge resurgence of the doublon state provides a clear and
experimentally-accessible dynamical manifestation of the
boundary-dependent energy spectrum of the two-particle
non-Hermitian Hubbard model.}

\section{Conclusions}
Non-Hermitian systems exhibit a variety
of intriguing phenomena that lack Hermitian counterparts, such as the 
strong dependence of the energy spectrum on the boundary conditions, the anomalous accumulation of bulk modes at the edges (non-Hermitian skin effect),
non-trivial point-gap topologies, and a wide variety of exotic bulk and edge dynamical phenomena. The physics of strongly-correlated systems described by effective non-Hermitian Hamiltonians are attracting a great interest recently, owing to the unique 
and largely unexplored properties arising from the interplay between particle correlations and non-Hermitian skin effect, such as non-Hermitian many-body localization and a nontrivial many-body topology. The interacting Hatano-Nelson model provides one of the simplest non-Hermitian extensions of the Hubbard model, which is a cornerstone theoretical model in condensed matter physics of strongly-correlated systems. In this work we presented exact analytical results of the spectral structure of the interacting Hatano-Nelson model in the two-particle sector of Hilbert space under different boundary conditions {\color{black} and unravelled emerging
 spectral and dynamical effects arising from the interplay
between particle correlation and non-Hermitian
terms in the Hamiltonian. These include (i) a novel spectral
phase transition of the Mott-Hubbard band on the
infinite lattice, from an open to a closed loop in complex
energy plane; (ii) a correlation-induced lifting of
degeneracy between spectra on the infinite and semi-infinite
lattices; (iii) dynamical dissociation of doublons,
i.e. instability of two-particle bound states, in the bulk of
the lattice; and (iv) a burst revival of the doublon state
when the two particles reach the lattice edge. The latter
phenomenon is a striking boundary-induced dynamical
phenomenon peculiar to non-Hermitian systems displaying
the non-Hermitian skin effect, and should provide an experimentally-accessible dynamical signature
of the boundary-dependent energy spectrum of the two-particle
non-Hermitian Hubbard model. Our results shed
new physical insights into the simplest non-Hermitian
extension of the Hubbard model in the few-particle
regime, highlight the strong impact of non-Hermiticity
on the spectral and dynamical features of two strongly-correlated
particles, and should stimulate further theroertical
and experimental studies on an emergent area of research.}

\medskip

% Acknowledgements
\medskip
\textbf{Acknowledgements} \par %delete if not applicable))
The author acknowledges the Spanish State Research Agency, through the Severo Ochoa
and Maria de Maeztu Program for Centers and Units of Excellence in R\&D (Grant No. MDM-2017-0711).

% References
\medskip

% Use the following code if you wish to generate your bibliography with BibTeX;
% replace the string "MSP-template" below with the name(s) of
% the BibTeX data base(s) you want to use.
% The resulting bibliography-output (the content of the .bbl file)
% must be pasted back into this file before submission.
% Please also include your BibTeX data base file(s) in your submission
% so that we can re-run BibTeX if necessary.
%
%\bibliographystyle{MSP}
%\bibliography{MSP-template}

% Figures/tables and captions
% Permission statements are required for all figures reproduced or adapted from previously published articles/sources. Please also ensure that all necessary permissions to reproduce images have been received
% Please remove these statements for original figures

% Please provide Biographies and photos for Essays, Feature Articles, Progress Reports, Reviews, and Perspectives for those authors who should be highlighted  
% These should be at most 100 words long
% For other article types this section can be removed
% Photographs should be 40mm broad and 50 mm high

%\begin{figure}
 % \includegraphics{bio-placeholder.jpg}
 % \caption*{Biography}
% \end{figure}

% Table of contents entry should be 50 - 60 words long
% Image should be 55 mm broad and 50 mm high or 110 mm broad and 20 mm high

  \appendix
  \renewcommand{\theequation}{A.\arabic{equation}}
\setcounter{equation}{0}
\small

\section{Energy spectrum on the infinite lattice}
The calculation of the eigenstates and corresponding energy spectrum on the infinite lattice entails to determine the eigenfunctions $\Phi_l$ of the difference equation [Eq.(15) in the main text]
\begin{equation}
E \Phi_l=-2\sigma J (\Phi_{l+1}+\Phi_{l-1})+U \Phi_l \delta_{l,0},
\end{equation}
with $\Phi_{-l}=\Phi_{l}$ and with the asymptotic condition that $|\Phi_l|$ is bounded as $l \rightarrow \infty$. In the above equation, we have set
\begin{equation}
\sigma= \cos(q-ih)
\end{equation}
which is parametrized by the real parameter $q=(k_1+k_2)/2$, corresponding to the quasi momentum of the particle center of mass. The spectral problem defined by Eq.(A1) with the appropriate boundary conditions corresponds to the well-known single-particle scattering problem on a lattice from a single impurity, but with a {\em complex} hopping amplitude $2 J \sigma$. The solutions to the single-impurity problem, bounded as $l \rightarrow \infty$, comprise two types of states.\\
(i) {\em Scattered states}, i.e. states that are bounded but do not vanish, i.e. oscillate, as $l \rightarrow \infty$. They are given by
\begin{equation}
\Phi_l=\left\{ 
\begin{array}{cc}
\sum_{\pm} \left( \frac{1}{2} \pm \frac{U}{8i \sigma J \sin Q}  \right) \exp( \pm i Q |l| ) & l \neq 0 \\
1 & l=0
\end{array}
\right.
\end{equation}
and the corresponding eigenenergies are
\begin{equation}
E_1=-4 \sigma J \cos Q,
\end{equation}
where $Q=(k_1-k_2)/2$ is an arbitrary real parameter, corresponding to the quasi momentum of the relative particle motion. Substitution of Eq.(A.2) into Eq.(A.4) yields
\begin{equation}
E_1=-4 J \cos Q \cos (q-ih)=-2J \cos(k_1-ih)-2J \cos(k_2-ih)
\end{equation}
which is the energy branch of two-particle scattered states given by Eq.(17) in the main text.\\
\\
(i) {\em Bound states}, i.e. states such that $\Phi_l \rightarrow 0$  as $l \rightarrow \infty$. It can be readily shown that the eigenstates corresponding to bound modees are given by
\begin{equation}
\Phi_l=\left\{ 
\begin{array}{cc}
\frac{U-E}{4 \sigma J \exp(-\mu)} \exp( - \mu |l| ) &  l\neq 0 \\
1 & l=0
\end{array}
\right.
\end{equation}
with corresponding eigenenergies
\begin{equation}
E_2=-2 \sigma J \cosh \mu.
\end{equation}
In the above equations, the complex parameter $\mu$ is given by
\begin{equation}
\mu= \log \left(-\frac{U}{4 \sigma J} \pm \sqrt{1+\left( \frac{U}{4 \sigma J}  \right)^2} \right) 
\end{equation}
where the sign on the right hand side of Eq.(A.8) should be chosen such that $\rm {Re} ( \mu)>0$.  Substitution of Eq.(A.8) into Eq.(A.7) yields
\begin{equation}
E_2=\sqrt{U^2+16 J^2 \sigma^2 }=\sqrt{U^2+16 J^2 \cos^2(q-ih) }
\end{equation}
which is the energy of the two-particle bound states given by Eq.(18) in the main text.

  \renewcommand{\theequation}{B.\arabic{equation}}
\setcounter{equation}{0}

\section{Energy spectrum on the semi-infinite lattice}
The spectral problem of the two-particle non-Hermitian Hubbard model on a semi-infinite lattice is defined by Eqs.(5) with the boundary conditions (10) and (11). In the spirit of the Bethe Ansatz \cite{r3,GDV}, generalized to account for the imaginary gauge field $h$, we may look for a solution to Eqs.(5) of the form
\begin{eqnarray}
u_{n,m} & = & A_1 \exp(ik_1 n+i k_2 m)+ A_2 \exp(-i k_1 n+ik_2 m-2 hn) \nonumber \\
& + & A_3 \exp(ik_1n-ik_2m-2hm) \nonumber \\
& + & A_4 \exp(-ik_1 n-i k_2 m-2hn-2hm) \nonumber \\
& + & A_5 \exp(ik_2 n+i k_1 m)+ A_6 \exp(-i k_2 n+ik_1 m-2 hn) \nonumber \\
& + & A_7 \exp(ik_2 n-i k_1 m-2hm) \nonumber \\
& + &  A_8 \exp(-i k_2 n-ik_1 m-2 hn-2hm)
 \end{eqnarray}
 for $n=1,2,3,...$ and $ 1 \leq m \leq n$, with $u_{n,m}=u_{m,n}$ for $m \geq n$ and
with the corresponding eigenenergy $E$ given by
\begin{equation}
E=-2J \cos(k_1-ih)-2J \cos (k_2-ih).
\end{equation}
In Eq.(B.1), the amplitudes $A_l$ should be determined by imposing the boundary condition $u_{n,0}=0$ for $n=1,2,3,...$ and the validity of Eqs.(5) for $n=m$, whereas the only constraint for the complex wave numbers $k_1$ and $k_2$ is the boundedness condition (11) of the wave function as $n,m \rightarrow + \infty$. To solve the spectral problem, we should distinguish two cases, that correspond in the Hermitian limit $h=0$ to the energy spectra of the two-particle scattered states ($E_1$ energy branch) and of the two-particle bound states ($E_2$ energy branch).\\
\\
{\it (i) $E_1$-energy branch.} In this case all amplitudes $A_l$ ($l=1,2,...,8$) in Eq.(B.1) are non-vanishing. After imposing the boundary condition $u_{n,0}=0$ for $n=1,2,3,...$ and the validity of Eqs.(5) for $n=m$, one obtains the following set of linear homogeneous equations for the vector $\mathbf{A}=(A_1,A_2,...,A_8)^T$ of the wave amplitudes
\begin{equation}
\mathcal{M} \mathbf{ A}=0
\end{equation}
where the non-vanishing elements of the $8 \times 8$ matrix $\mathcal{M}$ are given by
\begin{eqnarray}
\mathcal{M}_{11} =\mathcal{M}_{13}=\mathcal{M}_{22}=\mathcal{M}_{24}=\mathcal{M}_{35}=\mathcal{M}_{37}=\mathcal{M}_{46}=\mathcal{M}_{48}=1 \nonumber \\
\mathcal{M}_{51} = E-U+2J \exp(h+ik_1)+2J \exp(-h-ik_2) \;\;\;\;\;\; \nonumber \\
\mathcal{M}_{55} = E-U+2J \exp(h+ik_2)+2J \exp(-h-ik_1) \;\;\;\;\;\; \nonumber \\
\mathcal{M}_{62} = E-U+2J \exp(-h-ik_1)+2J \exp(-h-ik_2) \;\;\;\;\;\; \nonumber \\
\mathcal{M}_{67} = E-U+2J \exp(h+ik_2)+2J \exp(h+ik_1) \;\;\;\;\;\; \nonumber \\
\mathcal{M}_{73} = E-U+2J \exp(h+ik_1)+2J \exp(h+ik_2) = \mathcal{M}_{67}  \;\;\;\;\;\; \nonumber \\
\mathcal{M}_{76} = E-U+2J \exp(-h-ik_2)+2J \exp(-h-ik_1) = \mathcal{M}_{62}  \;\;\;\;\;\; \nonumber \\
\mathcal{M}_{84} = E-U+2J \exp(-h-ik_1)+2J \exp(h+ik_2) = \mathcal{M}_{55}  \;\;\;\;\;\; \nonumber \\
\mathcal{M}_{88} = E-U+2J \exp(-h-ik_2)+2J \exp(h+ik_1) = \mathcal{M}_{51}. \;\;\;\;\;\; \nonumber 
\end{eqnarray}
It can be readily shown that, for arbitrary complex wave numbers $k_{1,2}$, one has $\det \mathcal{M}=0$. This means that there exist non-trivial solutions to the spectral problem for arbitrary values of $k_{1,2}$, the only constraint being given by the boundedness condition (11). From Eq.(B.1), it readily follows that the wave function $u_{n,m}$ is bounded as $n,m \rightarrow + \infty$ provided that
\begin{equation}
0 \leq {\rm Im} (k_{1,2} ) \leq 2h.
\end{equation}
Therefore, the first branch $E_1$ of the energy spectrum is given by Eq.(B.2), where $k_{1}$ and  $k_{2}$ are arbitrary complex wave numbers satisfying the only constraint given by Eq.(B.4). Note that, if we restrict $k_1$ and $k_2$ to be real, i.e. ${\rm Im} (k_{1,2})=0$, the $E_1$-energy branch under SIBC reproduces the $E_1$-branch on the infinite lattice derived in Sec.3.2 and corresponding to scattered two-particle states. Remarkably, even though we allow the imaginary parts of $k_1$ and $k_2$ to be non vanishing and to vary in the range constrained by Eq.(B.4), the $E_1$-energy spectrum is not changed and describes the same area in complex energy plane as in the infinite lattice case (see the shaded dark areas of Figs.3 and 4). However, when the imaginary parts of $k_1$ and $k_2$ do not vanish, the wave functions are squeezed toward the edge $n=m=1$, corresponding to the two-particle non-Hermitian skin effect. \\
\\
{\it (ii) $E_2$-energy branch.} A different type of solutions is found by assuming $A_2=A_4=A_5=A_7=0$, so that the plane-wave expansion (B.1) contains only four components $\mathbf{A}^{\prime}=(A_1,A_3,A_6,A_8)^T$. The boundedness condition for the wave function $u_{n,m}$ as $n,m \rightarrow + \infty$ is met provided that
\begin{equation}
0 \leq {\rm Im}(k_1) \leq 2h \;\; ,\;\;\; - {\rm Im}(k_1) \leq  {\rm Im}(k_2) \leq 2h.
\end{equation}
Unlike the previous case, in order for the homogeneous system Eq.(B.3) to have non-trivial solutions it can be readily shown that the following solvability condition 
\begin{equation}
U=4iJ \sin \left( \frac{k_1-k_2}{2} \right) \cos \left( \frac{k_1+k_2}{2}-ih \right) 
\end{equation}
should be satisfied. The solvability condition (B.6) basically indicates that the two variables $Q \equiv (k_1-k_2)/2$ and $q \equiv (k_1+k_2)/2$ cannot be chosen independently one another: once one of the two variables has been arbitrarily chosen, with the only constraint imposed by the boundedness of the wave function $u_{n,m}$ at infinity [Eq.(B.5)], the other one is determined via Eq.(B.6). Let us assume 
\begin{equation}
q \equiv \frac{k_1+k_2}{2} 
\end{equation}
as the independent variable. The boundedness condition (B.5) yields the constraint $0 \leq {\rm Im} (q) \leq  2h$. The corresponding eigenenergy of the $E_2$-branch is given by Eq.(B.2), which using Eq.(B.6) can be written in the following form
\begin{eqnarray}
E_2 & =- & 2J \cos(k_1-ih)-2J \cos (k_2-ih) \nonumber \\
& = & -4J \cos (q-ih) \cos(Q) = 4J \cos (q-ih) \sqrt{1-\sin^2 Q} \nonumber \\ 
& = & \sqrt{U^2+16 J^2 \cos^2 (q-ih)}
\end{eqnarray}
Note that Eq.(B.8) is formally the same as Eq.(A.9) describing the energy dispersion curve of the Mott-Hubbard band on the infinite crystal, however while in the latter case the wave number $q$ must be real under SIBC $q$ is a complex number with the minimal constraint $0 \leq {\rm Im} (q) \leq  2h$.

\end{document}